\documentclass[conference]{IEEEtran}
\IEEEoverridecommandlockouts

\newtheorem{proof}{Proof}
\makeatletter
\newcommand{\Rmnum}[1]{\expandafter\@slowromancap\romannumeral #1@}
\makeatother

\usepackage{booktabs}
\usepackage{amsmath, amssymb, amsfonts}
\usepackage{graphicx}
\usepackage{bm}
\usepackage{multicol}
\usepackage{setspace}
\usepackage{subfigure}
\usepackage{epstopdf}
\usepackage{fancyhdr} 
\usepackage{indentfirst}
\usepackage{enumerate}
\usepackage{stfloats}
\usepackage{bm}
\usepackage{multirow}
\usepackage{threeparttable}
\usepackage{CJK}
\usepackage[justification=centering]{caption}
\usepackage{makecell}
\usepackage{caption}
\usepackage{cases}
\usepackage{textcomp}
\usepackage{algpseudocode}
\usepackage{amsmath}
\usepackage{graphics}
\usepackage{epsfig}
\usepackage{cite}
\usepackage{color}
\usepackage{caption}

\usepackage{geometry}
\begin{document}
\captionsetup[figure]{labelformat={default},labelsep=period,name={Fig.}}
\title{ 
	VQ-VAE Based Digital Semantic Communication with Importance-Aware OFDM Transmission
}

\author{\IEEEauthorblockN{Ming Lyu, Hao Chen, Dan Wang, Chen Qiu, Guangyin Feng, Nan Ma and Xiaodong Xu
		}
\thanks{
This work was supported by the National Science and Technology Major Project - Mobile Information Networks under Grant No. 2024ZD1300700.

M. Lyu is with the Department of Broadband Communication, Pengcheng Laboratory, Shenzhen, China, 518055, and also with the School of Future Technology, South China University of Technology, Guangzhou, China, 511442. Email: lvm@pcl.ac.cn.

H. Chen, D. Wang and C. Qiu are with the Department of Broadband Communications, Pengcheng Laboratory, Shenzhen, China, 518055, Emails: chenh03@pcl.ac.cn, wangd01@pcl.ac.cn, qiuch@pcl.ac.cn. 

G. Feng is with the School of Microelectronics, South China University of Technology, Guangzhou, China, 511442. Email: gyfeng88@scut.edu.cn.

N. Ma and X. Xu are with the State Key Laboratory of Networking and Switching Technology, Beijing University of Posts and Telecommunications, Beijing, China, 100876, and also with the Department of Broadband Communication, Pengcheng Laboratory, Shenzhen, China, 518055. Email: manan@bupt.edu.cn, xuxiaodong@bupt.edu.cn.
}
}
%\thanks{fjvfjvjb
%}
\maketitle

\begin{abstract}
Semantic communication (SemCom) significantly reduces redundant data and improves transmission efficiency by extracting the latent features of information. However, most of the conventional deep learning-based SemCom systems focus on analog transmission and lack in compatibility with practical digital communications. This paper proposes a vector quantized-variational autoencoder (VQ-VAE) based digital SemCom system that directly transmits the semantic features and incorporates the importance-aware orthogonal frequency division multiplexing (OFDM) transmission to enhance the SemCom performance, where the VQ-VAE generates a discrete codebook shared between the transmitter and receiver. At transmitter, the latent semantic features are firstly extracted by VQ-VAE, and then the shared codebook is adopted to match these features, which are subsequently transformed into a discrete version to adapt the digital transmission. To protect the semantic information, an importance-aware OFDM transmission strategy is proposed to allocate the key features near the OFDM reference signals, where the feature importance is derived from the gradient-based method. At the receiver, the features are rematched with the shared codebook to further correct errors. Finally, experimental results demonstrate that our proposed scheme outperforms the conventional DeepSC and achieves better reconstruction performance under low SNR region.
\end{abstract}

\begin{IEEEkeywords}
Semantic communication (SemCom), feature importance, vector quantised-variational autoencoder (VQ-VAE), orthogonal frequency division multiplexing (OFDM), reference signals.
\end{IEEEkeywords}

\section{Introduction}  
With the exponential growth of wireless data traffic driven by rapid advancements in information technology, conventional communication systems are reaching their physical and operational limits, highlighting the urgent need for more intelligent and efficient communication paradigms. Concurrently, the flourishing development of deep learning has inspired researchers to explore new paradigms for intelligent communication. Semantic communication (SemCom), which aims to extract latent information via deep learning, is regarded to significantly reduce the redundant data \cite{lu_semantics-empowered_2024, sun_semantic_2024}. However, most of the existing works transmit semantic features directly or map them to analog symbols, resulting in incompatibility with practical digital systems \cite{bourtsoulatze_deep_2019, sun_deep_2022,xie_deep_2021}.

To address the above mentioned challenges, the authors in \cite{huang_d-jscc_2025} extended the classical deep joint source-channel coding by quantizing semantic features which represented as vectors into scalars for digital transmission. In contrast, the vector quantized-variational autoencoder (VQ-VAE) is employed to generate discrete feature indices, which is suitable for digital transmission \cite{hu_robust_2023}. Moreover, VQ-VAE maps continuous latent variables to a discrete codebook, serving as a shared knowledge base between the transmitter and the receiver \cite{ye_codebook-enabled_2024}. By transmitting only the indices of discrete features within the codebook, the amount of data required to be transmitted is significantly reduced \cite{ma_task-oriented_2023}. However, transmission errors of feature indices lead to incorrect mappings in the codebook in complex channel environments, where the retrieved features differ substantially from the original ones, causing poor performance for image reconstruction. 

On the other hand, semantic importance has been widely investigated to enhance transmission efficiency and robustness by prioritizing the transmission of critical semantic features \cite{sun_deep_2022,zhou_feature_2024}. The authors in \cite{zhou_feature_2024} studied an importance-aware image transmission model in orthogonal frequency division multiplexing (OFDM)-based SemCom systems, which maps the critical semantic features into subcarriers with higher signal-to-noise ratio (SNR) to preserve key semantic information, thereby enhancing the image reconstruction accuracy. However, most of the existing OFDM-based SemCom systems still rely on end-to-end optimization models and lack error correction mechanisms of semantic features.
 
In this paper, we propose a novel VQ-VAE based digital SemCom system that directly transmits the semantic features and integrates the importance-aware OFDM transmission to maintain compatibility with practical digital systems and reduce transmission errors under complex channel environments, where the VQ-VAE generates a discrete codebook shared between the transmitter and receiver. At VQ-VAE encoder, the latent features are extracted and matched with the shared codebook to produce discrete features, which are then directly transmitted. To the best of our knowledge, most VQ-VAE-based studies focus on the transmission of semantic indices, while few have explored the direct transmission of semantic features, particularly in the context of OFDM transmission. Then, to enhance the protection of semantic features, the most important semantic features are allocated near the reference signals during the OFDM transmission process, where the feature importance is determined through gradient-based evaluation. Finally, by leveraging a shared codebook and a feature rematching mechanism at the receiver, the system enables effective error correction during the wireless transmission of semantic features.

\section{System Model} 
\begin{figure*}[htbp]
	\centering
	\includegraphics[width=0.92\linewidth]{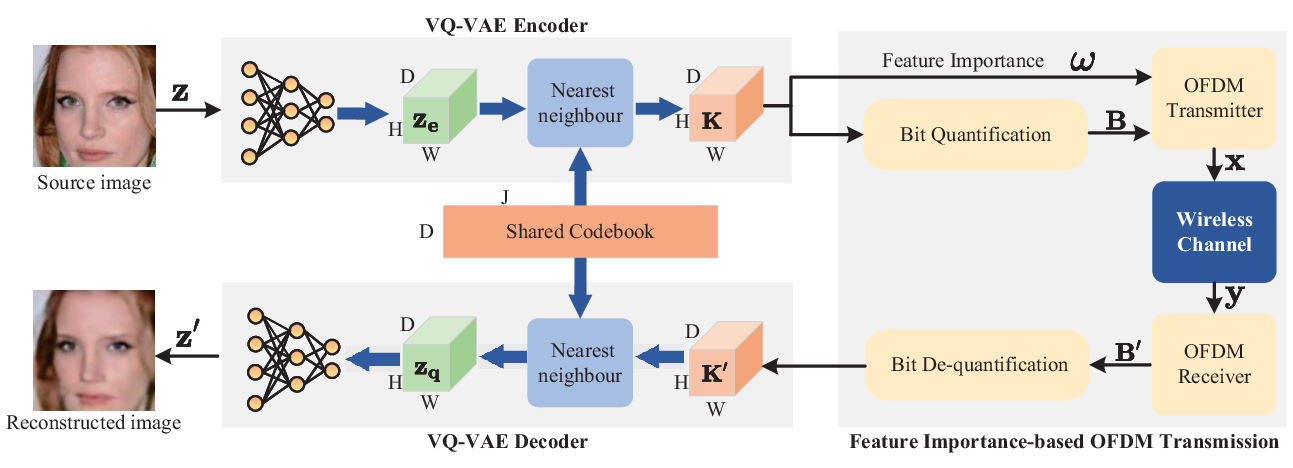}
	\caption{VQ-VAE based digital SemCom system with importance-aware OFDM transmission.}
	\label{VQVAE}
\end{figure*}
As illustrated in Fig. \ref{VQVAE}, we consider a VQ-VAE based digital SemCom system with importance-aware OFDM transmission, where the transmitter and receiver are equipped with the VQ-VAE encoder and decoder, respectively. The VQ-VAE encoder and decoder consist of a CNN for feature extraction and a nearest-neighbor operation for feature discretization. The CNN is composed of a series of convolutional layers followed by parametric ReLU activation functions and residual blocks. A common codebook $\mathcal{E} =[\mathbf{e}_1, \dots \mathbf{e}_j, \dots \mathbf{e}_J ]$ is generated by the VQ-VAE, which is shared by the transmitter and receiver, with  $\mathbf{e} _j \in \mathbb{R} ^{D}$ being fixed after the VQ-VAE training phase.

At transmitter, the VQ-VAE encoder is employed to extract discrete semantic features via the shared codebook. To protect the semantic features, a gradient-based evaluator is implemented to determine the importance weights of these features, which allows the bit-quantized features to be allocated at the OFDM transmitter. Here, the most important features are allocated near the reference signals to reduce the channel distortion. At receiver, channel estimation is usually more accurate for data located closer to the reference signals, resulting in lower errors for important features and enhancing transmission accuracy. After channel estimation and demodulation at the OFDM receiver, the features are reconstructed via inverse bit-quantization and subsequently input into the VQ-VAE decoder. Finally, the received features are rematched with the shared codebook to correct errors. Next, the detail of the proposed semantic transmission scheme is introduced in sequel.

\subsection{VQ-VAE Encoder}
At the VQ-VAE encoder, an input image is represented by a vector $\mathbf{z} \in \mathbb{R} ^l$, and the neural networks of VQ-VAE first encodes $\mathbf{z}$ into a feature tensor $\mathbf{z_e} \in \mathbb{R} ^{D\times H\times W}$, where $D$ is the number of features, and $H\times W$ is the shape of each feature. The process of extracting semantic features is represented as
\begin{align}
	\label{2.1} 
	\mathbf{z_e}=N_{ \boldsymbol{\theta}_1 }(\mathbf{z} ) , 
\end{align}
where $N_{ \boldsymbol{\theta}_1 }(\cdot )$ denotes the encoder CNN with parameter $\boldsymbol{\theta}_1$. 

Then, the discrete latent features $\mathbf{K}=[\mathbf{k} _1,\dots \mathbf{k} _i,\dots ,\mathbf{k} _D]$ where $\mathbf{K}\in \mathbb{R} ^{D\times H\times W}$ are calculated by a nearest neighbor look-up using the shared codebook $\mathcal{E}$ with the feature tensor $\mathbf{z_e}$. Specifically, each vector in $\mathbf{z_e}$ is replaced by the closest codebook entry in $\mathcal{E}$ in terms of Euclidean distance, i.e., \cite{van_den_oord_neural_2017}
\begin{align}
	\label{3.1} 
	\mathbf{K}=\arg \min _{\mathbf{e}_{j}}\left\|\mathbf{z_{e}}-\mathbf{e}_{j}\right\|_{2}, \forall \mathbf{e}_{j} , 
\end{align}
where $\left\| \cdot  \right\|_{2}$ is the 2-norm, the resulting features $\mathbf{K}$ thus consist of selected codebook vectors. Notably, we transmit the features $\mathbf{K}$ rather than their corresponding indices in the codebook.

Subsequently, $\mathbf{K}$ is input into the OFDM-based digital semantic transmission module, which transmits the bit-quantized semantic data according to their importance. The detailed description of this module will be provided in Section III.

\begin{figure*}[htbp]
	\centering
	\includegraphics[width=1\linewidth]{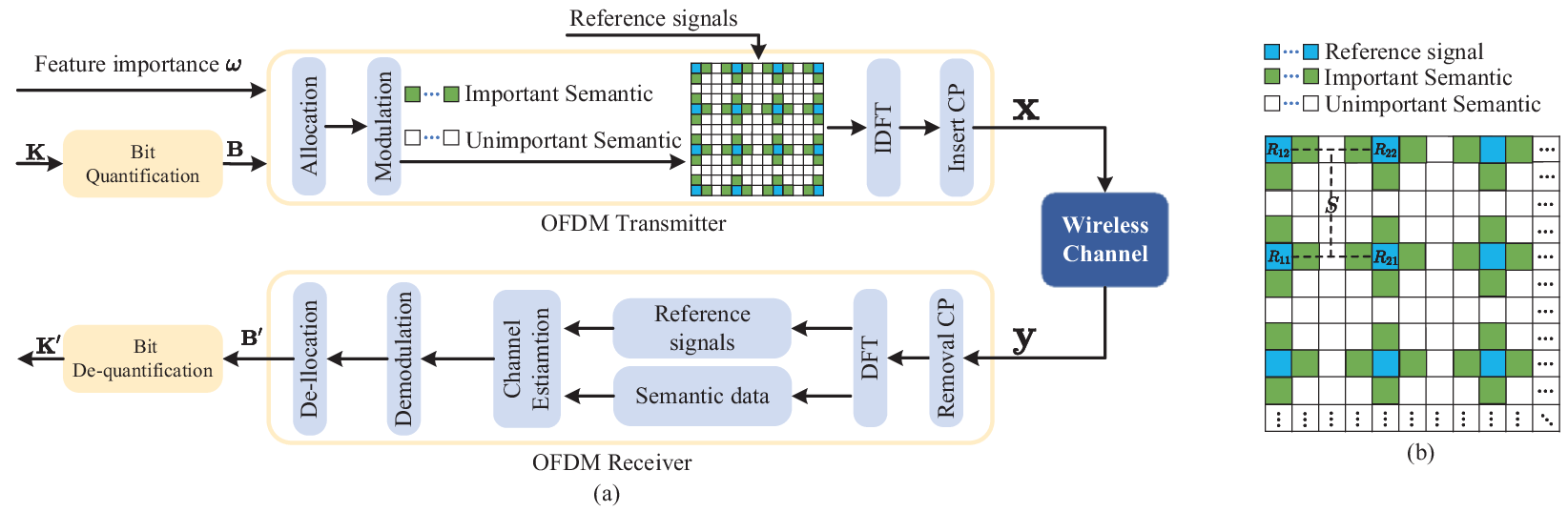}
	\caption{(a) The proposed OFDM transmission with feature importance framework; (b) REs distribution on the OFDM time-frequency resources grid.}
	\label{DiSeT}
\end{figure*}

\subsection{VQ-VAE Decoder}
After transmission, the features ${\mathbf{K}}'=[\mathbf{k}' _1,\dots \mathbf{k}' _i,\dots ,\mathbf{k}' _D]$ where ${\mathbf{K}}' \in \mathbb{R} ^{D\times H\times W}$ are inputed
into decoder, where the $i$-th feature error $E_i$ can be represented as
\begin{align}
	\label{3.3} 
	E_i = \left\| \mathbf{k}_i -  {\mathbf{k}_i }' \right\|_{2} = E_{b,i} + E_{h,i},
\end{align}
where $E_{h,i}$ and $E_{b,i}$ denote the wireless channel transmission errors and the bit quantization errors of the $i$-th feature, respectively.

Then, the distorted elements in ${\mathbf{K}}'$ are rematched to the vector $\mathbf{e}_{j}$ in the shared codebook by a nearest neighbor look-up to correct the errors, i.e.,
\begin{align}
	\label{3.2} 
	\mathbf{z_q}=\arg \min _{\mathbf{e}_{j}}\left\|{\mathbf{K}}'-\mathbf{e}_{j}\right\|_{2}, \forall \mathbf{e}_{j} ,
\end{align}
where $\mathbf{z_q}\in \mathbb{R} ^{D\times H\times W}$ is the corrected semantic features. 

Next, we further analyze the error correction process of the rematching mechanism. When the semantic features being transmitted in $\mathbf{K}$ corresponds to the basis vector $\mathbf{e}_{i} $ in the codebook, the minimum distance $d_{min}$ between $\mathbf{e}_{i}$ and other vectors $\mathbf{e}_{j}$ in the codebook can be calculated as
\begin{align}
	\label{3.4} 
	d_{min} =\min _{j=1}^{J}\left\|\mathbf{e}_{j}-\mathbf{e}_{i}\right\|_{2}.
\end{align}
When $E_i< \frac{d_{min}}{2} $, the transmission error of the feature can be perfect corrected by rematching with the vectors in the shared codebook through a nearest-neighbor look-up at the receiver. Additionally, when $E_i> \frac{d_{min}}{2} $, the error of feature mapped to a neighboring vector, allowing for imperfect but effective correction. For the index-based transmission method, features are directly retrieved from the codebook using the transmitted indices, making it unable to correct transmission errors. In addition, an erroneous index may result in the mapping to a vector that is significantly distant from the original one. 

Finally, the recovered semantic features $\mathbf{z_q}$ are input into the VQ-VAE decoder to reconstruct image $\mathbf{{\mathbf{z} }' }$, which can be represented as
\begin{align}
	\label{2.7} 
	\mathbf{{\mathbf{z} }' } =N_{ \boldsymbol{\theta}_2  }^{-1}(\mathbf{{\mathbf{K}}'  } ) , 
\end{align}
where $N_{ \boldsymbol{\theta}_2 }^{-1}(\cdot )$ denotes the decoder CNN with parameter $\boldsymbol{\theta}_2$. The decoder CNN inverts the operations performed by the encoder through a series of transpose convolutional layers. Our proposed feature rematching mechanism based on VQ-VAE significantly improves the robustness of semantic feature transmission and the quality of image reconstruction.

\section{OFDM Transmission with Feature Importance Awareness} 
In this section, we propose a OFDM transmission with feature importance framework, as shown in Fig. \ref{DiSeT}. Due to the non-uniform channel quality across the OFDM time-frequency resources, we adopt a discrete reference signals distribution pattern \cite{lyu_reference_2025}, placing important semantic signals around the reference signals.

\subsection{OFDM Transmitter} 
After the VQ-VAE encoder, we extract the continuous-valued semantic features $\mathbf{k} _i$ undergo uniform bit-quantization \cite{zhang_deep_2023}, yielding a discrete bitstream $\mathbf{b} _i$, which can be represented as
\begin{align}
	\label{2.2} 
	\mathbf{b} _i= Q_c(\mathbf{k} _i) , 
\end{align}
where $Q_c(\cdot )$ denotes the uniform bit-quantizer, $c$ is the number of quantization bits. The quantization results can be represented as $\mathbf{B}  = [\mathbf{b} _1,\dots \mathbf{b} _i,\dots,\mathbf{b} _D]$ with $\mathbf{b} _i \in \left \{ 0, 1 \right \} ^L$, where $L$ denotes the total number of each feature bits. Since different semantic features $\mathbf{b} _i$ contribute unequally to image reconstruction fidelity, a gradient-based evaluator is adopted to calculate the importance of each feature.

The impact of each feature on image reconstruction is evaluated by analyzing the gradient variations of the loss $L (\mathbf{z} ,\mathbf{z}^\prime )$ within the neural network, i.e.,
\begin{align}
	\label{4.0} 
	L (\mathbf{z} ,\mathbf{z}^\prime ) = \frac{1}{l} \left \| \mathbf{z}^\prime-\mathbf{z}  \right \| ^2,
\end{align}
where $\left\| \cdot  \right\|^{2}$ denotes the squared Euclidean norm.

Specifically, we first compute the gradients with respect to the $i$-th feature $ \mathbf{k} _i$ and obtain a gradient matrix $\frac{\partial L (\mathbf{z} ,\mathbf{z}^\prime )}{\partial \mathbf{k} _i} $. Then, by applying global average pooling to the gradient matrix, the importance weights vector of the feature tensor $\mathbf{K}$ is denoted as $\bm{\omega}=[\omega _1,\dots \omega _i,\dots ,\omega _D]$, i.e., \cite{zhou_feature_2024}
\begin{align}
	\label{4.1} 
	\omega _i = \frac{1}{HW} \sum_{n=1}^{H} \sum_{m=1}^{W}   \frac{\partial L (\mathbf{z} ,\mathbf{z}^\prime )}{\partial k_{i,nm}}.
\end{align}

For different values of $\omega _i$, the corresponding semantic features are mapped onto OFDM time-frequency resource elements (REs) with varying priorities, as shown in Fig. \ref{DiSeT}(b). For simplicity, each unit in the time-frequency grid is denoted as a RE: blue REs represent the reference signals, green REs represent the important semantic symbols, blank REs represent the unimportant semantic symbols. Subsequently, we map each part step by step onto the OFDM time-frequency grid.

Firstly, the discrete reference signals are uniformly distributed over the time and frequency domains, which facilitates accurate estimation of the channel's time-varying and frequency-selective fading. Then, the numbers of important and unimportant semantic features are calculated. Since semantic symbols transmitted closer to reference signals achieve higher accuracy, four green REs are allocated around each reference signal, while the remaining REs are allocated to unimportant semantic features. Suppose the total number of REs within a coherent processing interval (CPI) is $N_{CPI}$, and the total number of reference signals is $N_{ref}$. Then, the number of REs allocated to important semantic is $4N_{ref}$, and the number of unimportant semantic is $N_{CPI}-5N_{ref}$. Accordingly, the total number of important semantic features is denoted as $D_{imp} = D\frac{4N_{ref}}{N_{CPI}-N_{ref}} $. Finally, the top-$D_{imp}$ semantic features, ranked by feature importance $\bm{\omega}$, are modulated and sequentially mapped onto the green REs, while the remaining semantic features are modulated and mapped onto the remaining blank REs.

After the semantic symbols are mapped onto the time-frequency resources, an inverse discrete Fourier transform (IDFT) is applied and a cyclic prefix (CP) is added to generate the transmitted signal $\mathbf{x} \in \mathbb{R} ^{N_{ c}\times N_{ t}}$, as shown in Fig. \ref{DiSeT}(a). The signal $\mathbf{x}$ then propagates through a time-frequency dual-varying channel, resulting in the received signal $\mathbf{y} \in \mathbb{R} ^{N_{ c}\times N_{ t}}$ expressed as
\begin{align}
	\label{2.5} 
	\mathbf{y}  = \mathbf{h} *\mathbf{x}  + \mathbf{n}  , 
\end{align}
where $*$ denotes convolution operation, $\mathbf{h} \in \mathbb{R} ^{N_{ c}\times N_{ t}}$ is the impulse response of the multipath channel, and $\mathbf{n}$ is the circularly symmetric complex Gaussian noise with zero mean and variance $\mathcal{CN} (\mathbf{0},\sigma^{2}\mathbf{I} )$, where $\mathbf{I} \in \mathbb{R} ^{N_{ c}\times N_{ t}}$ being the identity matrix.

\subsection{OFDM Receiver} 
At the OFDM receiver, the received signal $\mathbf{y}$ undergoes CP removal, followed by a discrete Fourier transform (DFT) to convert it into the frequency domain, thereby enabling the separation of reference signals and semantic data.

Subsequently, channel estimation is performed for the semantic data. The least squares (LS) method is first employed to estimate the channel $\mathbf{h}_{LS}$ corresponding to the reference signals. Based on $\mathbf{h}_{LS}$, bilinear interpolation is then applied to obtain the estimated channel $\mathbf{\hat{h}}$ for the semantic data. Finally, the estimated channel $\mathbf{\hat{h}}$ is used to equalize the semantic data.

In the following, the transmission errors in the feature importance-based OFDM transmission are analyzed through the channel estimation procedure. The channel estimation error $E_h$, primarily arises from the LS estimation and bilinear interpolation processes, can be expressed as
\begin{align}
	\label{4.2} 
	E_h=\mathbf{h}-\mathbf{\hat{h} } = E_{LS}+E_{sd},
\end{align}
where $E_{LS}$ and $E_{sd}$ denotes the error brought by the LS estimation and interpolation processes, respectively. The estimation error $E_{LS}$ at the reference signal positions can be expressed as
\begin{align}
	\label{4.3} 
	E_{LS}=\mathbb{E} \left \{ \left \| \mathbf{\mathbf{h}}_{LS}-\mathbf{h } \right \|^2  \right \}   = \mathbb{E} \left \{ \mathbf{n}^\mathrm{H} (\mathbf{x}\mathbf{x}^\mathrm{H})^{-1} \mathbf{n} \right \}=\frac{\sigma^{2}}{\sigma_x^{2}},
\end{align}
where $\sigma_x^{2}$ denotes power of the reference signals.

Suppose the values of the reference signals $R_{11}(n_0,m_0)$, $R_{12}(n_0,m_1)$, $R_{21}(n_1,m_0)$ and $R_{22}(n_1,m_1)$ have been obtained through LS estimation, as shown in Fig. \ref{DiSeT}(b). Interpolation is first applied along the horizontal axis, and then along the vertical axis to compute the value of semantic data $S(n,m)$. The linear interpolation error $E_{sd}$ is represented by the Lagrange interpolation remainder term, which can be expressed as
\begin{align}
	\nonumber
	& E_{sd} \le  \frac{M_0}{2} (m-m_0) (m_1-m) \\ \label{Esd1}
	&+ \frac{[M_1(m_1-m)+M_2(m-m_{0})](n-n_0) (n_1-n)}{2(m_{1}-m_{0})},
\end{align}
where $ M_0, M_1, M_2$ are the bounded of the second-order partial derivatives of the interpolated function.
\begin{proof} 
	Please see Appendix \ref{secA}.  
\end{proof}

%\begin{proof} 
%	Please see Appendix \ref{secA}.  
%\end{proof}
%\begin{align}
%	\label{4.4} 
%	E_{sd}&  \approx \frac{f  ^{\prime \prime}(\xi ,\eta )}{2} (m-m_0) (m_1-m) \nonumber \\
%	&+\frac{f  ^{\prime \prime}(\xi ,m_{1} )}{2} \frac{(m-m_{0})(n-n_0) (n_1-n)}{m_{1}-m_{0}}  \nonumber  \\
%	&+ \frac{f  ^{\prime \prime}(\xi ,m_{0})}{2} \frac{(m_1-m)(n-n_0) (n_1-n)}{m_{1}-m_{0}}
%\end{align}
%where $\xi$ and $\eta$ are values between $[n_{0},n_{1}]$ and $[m_{0},m_{1}]$ respectively, these values depend on the relative positions of the semantic data and the reference signals. 

From (\ref{Esd1}), it is observed that the distance between the semantic data and the reference signals controls the interpolation error $E_{sd}$. When important semantic data are positioned close to the reference signals, their errors approximate the reference signals estimation error $E_{LS}$. In contrast, the unimportant semantic data error is further affected by interpolation errors $E_{sd}$, which depend on the density of the reference signals. A higher reference signals density results in a shorter distance to the semantic data and thus a smaller interpolation error $E_{sd}$.

After channel equalization, the semantic symbols are demodulated and the original feature ordering is restored to obtain the output $\mathbf{B ^\prime } $. The bitstream $\mathbf{B ^\prime }$ is then dequantized to recover the semantic features ${\mathbf{K}}'$, which can be represented by
\begin{align}
	\label{2.7} 
	{\mathbf{k} _i ^\prime} =Q_c^{-1}({\mathbf{b} _i ^\prime}) , 
\end{align}
where $Q_c^{-1}(\cdot)$ denotes the uniform bit-dequantizer. ${\mathbf{K}}'$ is input into the VQ-VAE decoder, where further error correction is performed and the image is reconstructed.

Our proposed OFDM-based feature importance transmission (FIT) scheme performs prioritized resource allocation through channel conditions, where high-criticality semantic features are allocated to time-frequency resources exhibiting superior transmission characteristics, while low-importance features are mapped to channel disadvantaged resources. This design enhances the protection of critical semantic information, thereby improving the overall performance of SemCom.

\section{Simulations and Numerical Results}   %5
In this section, we verify the effectiveness of the proposed OFDM-based digital SemCom system by numerical results.

\subsection{Simulation Setup}
1) \textit{Dataset:} Given our focus on the image reconstruction, we have employed the CelebA-HQ dataset \cite{karras_progressive_2018} containing 30,000 high-quality celebrity faces and resampled to 128×128 pixels for convenience of testing. 

2) \textit{OFDM Transceiver:} Here, we consider the Rayleigh fading channel. The carrier frequency and bandwidth we considered are 2.4 GHz and 20MHz, and can be applied across various frequency ranges. For modulation, 16-QAM is applied to the bitstream, and the CP length is set to 72. In a CPI, the OFDM time-frequency resource grid is composed of 448 OFDM symbols and 792 subcarriers. The reference signals are uniformly spaced with intervals of 4 in the time domain and 6 in the frequency domain. For simplicity, this works considers a single antenna system to focus on the proposed digital SemCom system.

3) \textit{VQ-VAE Implementation Details:} The VQ-VAE encoder consists of two strided convolutional layers with stride 2 and window size 4×4, followed by two residual 3×3 blocks. The VQ-VAE decoder follows a similar structure. We use the ADAM optimiser with learning rate 2e-4 and the codebook size is 1024. All models were trained under noiseless conditions, using an RTX 3070Ti GPU. The VQ-VAE encoder compresses the original image with dimensions 3×128×128 into semantic features with dimensions 16×32×32, achieving a compression rate of 1/3.

Peak signal-to-noise ratio (PSNR) and structural similarity index measure (SSIM) are used to evaluate the quality of the reconstructed images. We compare the performance of the following methods:
\begin{itemize}
	\item Proposed VQ-VAE + FIT: The proposed VQ-VAE with importance-aware transmission method.
	\item Proposed VQ-VAE: The proposed VQ-VAE without distinguishing importance-aware transmission.
	\item VQ-VAE + FIT: The VQ-VAE does not rematch the codebook at the receiver, with importance-aware transmission method.
	\item VQ-VAE: The VQ-VAE does not rematch the codebook at the receiver, without distinguishing importance-aware transmission.
	\item DeepSC: The modified deep source coding scheme \cite{bourtsoulatze_deep_2019}.
\end{itemize}

\subsection{Performance Evaluation}
\begin{figure}[htbp]
	\centering
	\includegraphics[width=1\linewidth]{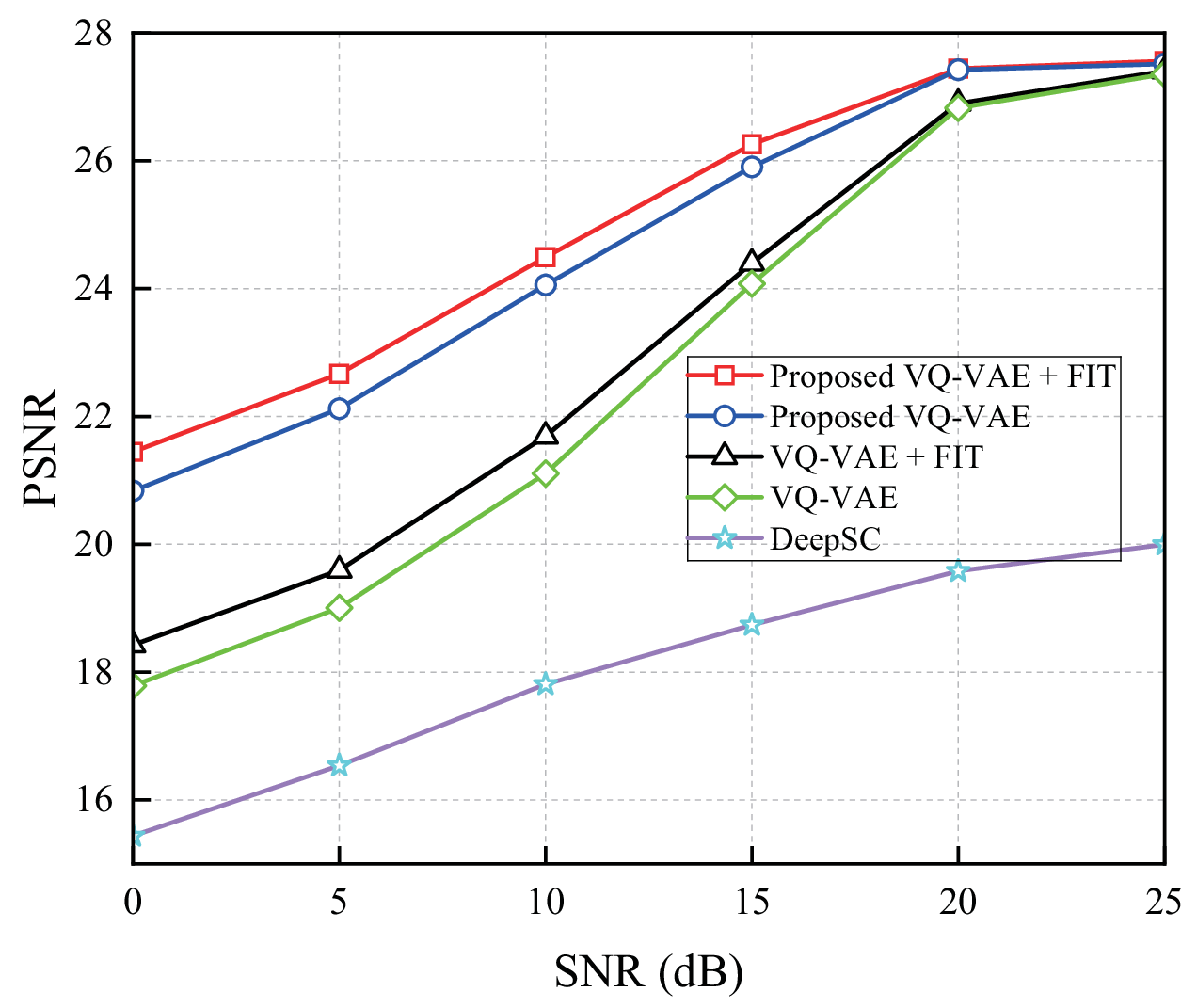}
	\caption{The PSNR of the different schemes versus SNR.}
	\label{psnr}
\end{figure}

\begin{figure}[htbp]
	\centering
	\includegraphics[width=1\linewidth]{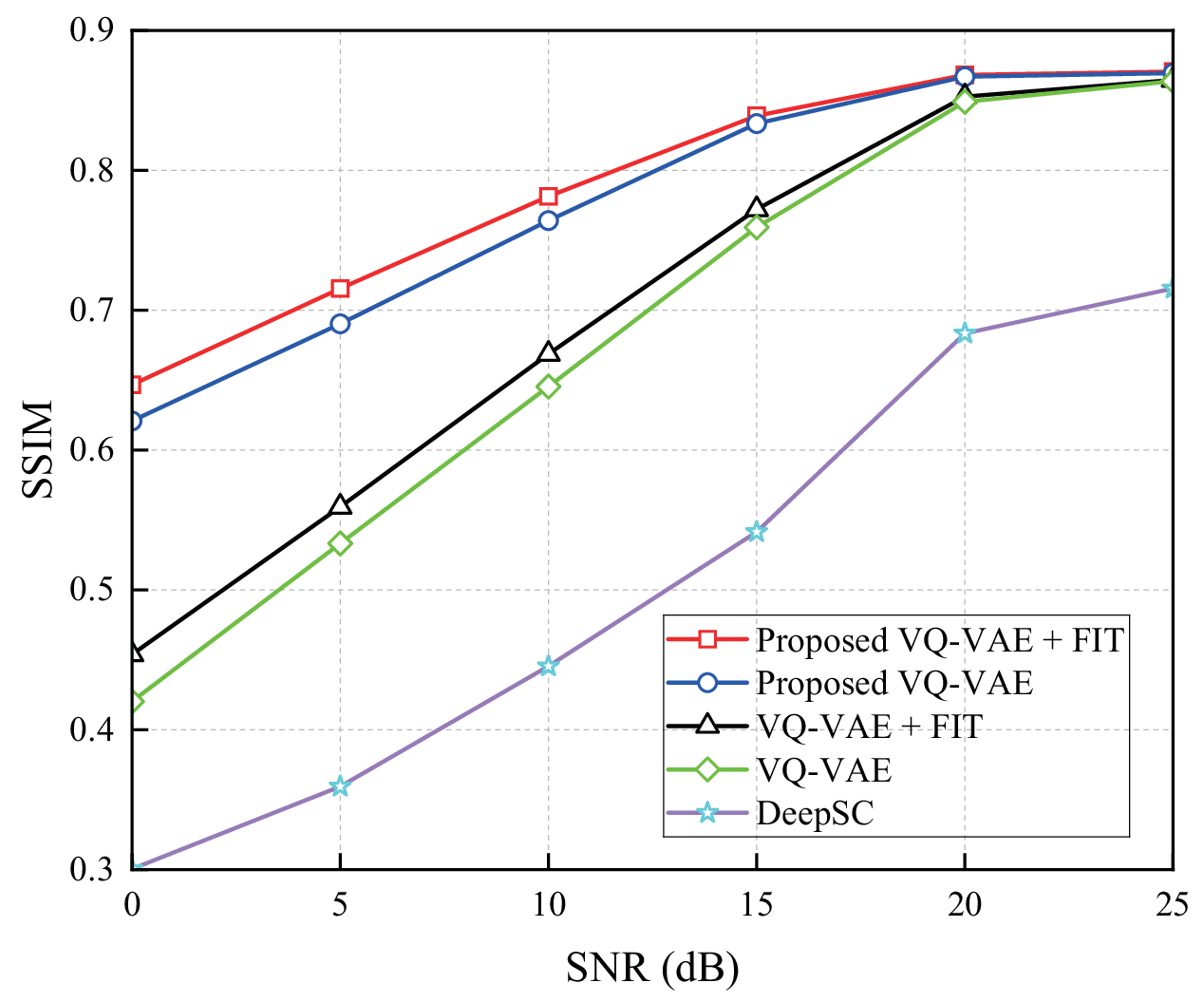}
	\caption{The SSIM of the different schemes versus SNR.}
	\label{ssim}
\end{figure}

Fig. \ref{psnr} and Fig. \ref{ssim} show the PSNR and SSIM versus SNR for different schemes. The performance increases with SNR for all schemes, and the proposed scheme consistently outperforms the baseline DeepSC scheme. Compared to conventional semantic encoder-decoder architectures, VQ-VAE introduces a vector quantization process that discretizes continuous features into codebook. This discretization improves noise tolerance, leading to superior performance compared to DeepSC. Moreover, the proposed VQ-VAE corrects semantic errors caused by transmission noise at the receiver by rematching with the codebook, further improving the performance of semantic transmission. Additionally, the OFDM transmission method based on semantic importance further enhances the performance of SemCom. This not only implies that SemCom is compatible with conventional communication frameworks, but also utilizes semantic characteristics to improve communication system performance.

Our proposed scheme demonstrates greater effectiveness under low SNR region. For example, at 15 dB, the proposed VQ-VAE + FIT scheme achieves a 10.5\% performance gain over the VQ-VAE scheme, and this gain further increases to 34.1\% at 5 dB, as shown in Fig. \ref{ssim}. In addition, the performance gains provided by the proposed VQ-VAE are more significant than semantic importance transmission method.

\begin{figure*}[htbp]
	\centering
	\includegraphics[width=1\linewidth]{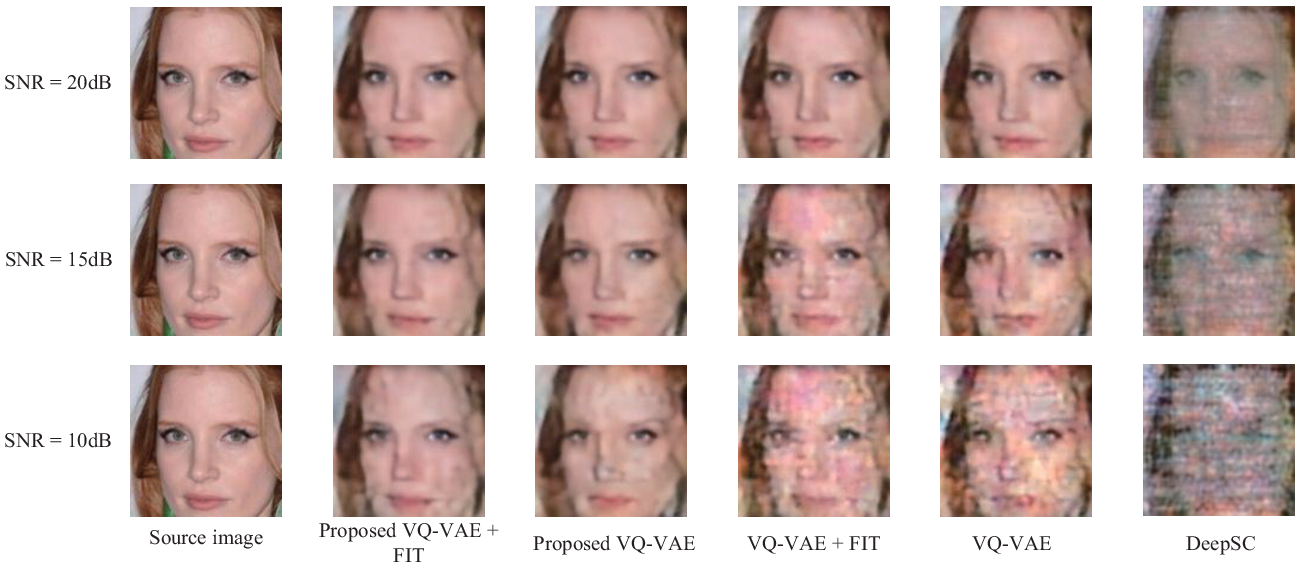}
	\caption{Visual reconstruction results under different semantic transmission schemes versus SNR.}
	\label{image}
\end{figure*}

Fig. \ref{image} presents the visual reconstruction results under different semantic transmission schemes versus SNR. The proposed VQ-VAE schemes clearly outperform the DeepSC, primarily due to the noise robustness provided by discretized semantic features. Moreover, our proposed VQ-VAE scheme demonstrates substantial improvements in image reconstruction quality, particularly under low SNR region. By reducing the transmission error in important semantic features, the key details and textures of the image are significantly improved.

\section{Conclusion}     %6
In this paper, we presented a novel VQ-VAE based digital SemCom system with importance-aware OFDM transmission. Firstly, a discrete codebook was generated by VQ-VAE and shared between the transmitter and receiver. The VQ-VAE encoder extracted the latent features and then matched with the shared codebook to produce discrete features for digital transmission. Then, to protect the semantic information, an importance-aware OFDM transmission strategy was proposed, which allocated the critical semantic signals near the reference symbols in the OFDM time-frequency grid based on the feature importance derived from the gradient-based method. Finally, experimental results demonstrated the superiority of our proposed SemCom system over the conventional DeepSC in reconstruction performance, with particularly notable improvements under low SNR region and strong compatibility with practical digital system. Future work will concentrate on end-to-end optimization of semantic transmission efficiency and resource allocation strategies.

% \section*{Acknowledgments}
% This should be a simple paragraph before the References to thank those individuals and institutions who have supported your work on this article.

%\section*{Acknowledgment}
%This work was supported by the

% \section*{References}

\appendices 

\section{The bilinear interpolation error} \label{secA}    %6
The linear interpolation along the horizontal axis is first performed $S_{1}$ using $R_{11}$ and $R_{21}$, which is expressed as
\begin{align}  \label{s11}
	S_1(n,m_0) & \approx R_{11}+\frac{n-n_{0}}{n_{1}-n_{0}} (R_{21}-R_{11}), \\ 	\label{s1} 
	& =\psi  _n R_{21} +(1-\psi  _n )R_{11},
\end{align}
where (\ref{s1}) is obtained by substituting $\psi  _n = \frac{n-n_{0}}{n_{1}-n_{0}}$ into (\ref{s11}).

Similarly, the horizontal interpolation at $S_{2}$ using $R_{12}$ and $R_{22}$ is given by
\begin{align}  \label{s22}
	S_{2}(n,m_1) & \approx R_{12}+\frac{n-n_{0}}{n_{1}-n_{0}} (R_{22}-R_{12}), \\ 	\label{s2} 
	& =\psi  _n R_{22} +(1-\psi  _n )R_{12} ,
\end{align}
where (\ref{s2}) is obtained by substituting $\psi  _n = \frac{n-n_{0}}{n_{1}-n_{0}}$ into (\ref{s22}).

Finally, the vertical interpolation at $S$ using $S_{1}$ and $S_{2}$ is performed as
\begin{align}  \label{s33}
	S(n,m) &  \approx S_{1}+\frac{m-m_{0}}{m_{1}-m_{0}} (S_{2}-S_{1}),  \\ 	\label{s3} 
	& =\psi  _m S_{2}+(1-\psi  _m )S_{1} ,
\end{align}
where (\ref{s3}) is obtained by substituting $\psi  _m =\frac{m-m_{0}}{m_{1}-m_{0}}$ into (\ref{s33}).

By substituting (\ref{s1}) and (\ref{s2}) into (\ref{s3}), the result $S(n,m) $ of bilinear interpolation is rewritten as
\begin{align}
	\nonumber
	S(n,m) &= \psi  _m[\psi  _n R_{22} +(1-\psi  _n )R_{12}] \\  \label{s4}
	&~~~~ +(1-\psi  _m )[\psi  _n R_{21} +(1-\psi  _n )R_{11}],   \\ \nonumber
	&=(1-\psi  _m )(1-\psi  _n )R_{11}+\psi  _m (1-\psi  _n )R_{12} \\ \label{s44}
	&~~~~+\psi  _n(1-\psi  _m ) R_{21}+\psi  _m \psi  _n R_{22},
\end{align}
where (\ref{s44}) is obtained by simplifying (\ref{s4}).

The bilinear interpolation error $E_{sd}$ accumulates from the remainders of the involved linear interpolations. When the $S_1(n,m_0)$ to be interpolated possesses a second-order derivative, i.e., $S_1\in C^2([n_0,n_1])$, the error of linear interpolation at any point $n\in[n_0,n_1]$ is given by
\begin{align}  \label{e11}
	R_1(n,m_0)&=S_1(n,m_0)-P_1(n,m_0), \\ 	\label{e1}   
	&=\frac{f  ^{\prime \prime}(\xi_1 ,m_0)}{2} (n-n_0) (n_1-n),   
\end{align}
where (\ref{e1}) is obtained by applying Taylor’s theorem with the Lagrange remainder to (\ref{e11}); $\xi_1\in[n_0,n_1]$; $R_1(n,m_0)$ is the interpolation remainder term; and $P_1(n,m_0)$ is the interpolation polynomial.

Similarly, the interpolation remainder at $S_2(n,m_1)$ is expressed as
\begin{align}
	\label{e2} 
	R_2(n,m_1)=\frac{f  ^{\prime \prime}(\xi_2 ,m_1)}{2} (n-n_0) (n_1-n),
\end{align}
where $\xi_2\in[n_0,n_1]$.

The final bilinear interpolation error is obtained by accumulating the remainders from the two horizontal linear interpolation steps with the vertical interpolation remainder. According to (\ref{s3}), the interpolation remainder at $S(n,m)$ is expressed as
\begin{align}
	\nonumber
	R(n,m)&=\frac{f  ^{\prime \prime}(\bar{\xi} ,\eta )}{2} (m-m_0) (m_1-m) \\ \label{ee}
	&+\psi  _m R_2+(1-\psi  _m )R_1,   \\  \nonumber
	&=\frac{f  ^{\prime \prime}(\bar{\xi} ,\eta )}{2} (m-m_0) (m_1-m) \\
	&+\frac{f  ^{\prime \prime}(\xi_2 ,m_1)}{2} \frac{(m-m_{0})(n-n_0) (n_1-n)}{m_{1}-m_{0}} \nonumber  \\ 	\label{e} 
	&+\frac{f  ^{\prime \prime}(\xi_1 ,m_0)}{2} \frac{(m_1-m)(n-n_0) (n_1-n)}{m_{1}-m_{0}},
\end{align}
where (\ref{e}) is obtained by substituting (\ref{e1}) and (\ref{e2}) into (\ref{ee}); $\eta\in[m_0,m_1]$; and $\bar{\xi}\in[n_0,n_1]$.

Since $\xi_1$, $\xi_2$ and $\eta$ are unknown, the interpolation error is typically assessed via upper bounds on the remainder term. To derive such bounds, we consider that the second-order partial derivatives of the interpolated function are bounded within the interpolation region; that is, there exist constants $ M_0, M_1, M_2 > 0$ such that
\begin{align}
	\label{M} 
	&|f  ^{\prime \prime}(\bar{\xi} ,\eta )| \le M_0, \\
	&|f  ^{\prime \prime}(\xi_1 ,m_0 )| \le M_1, \\ \label{MM} 
	&|f  ^{\prime \prime}(\xi_2 ,m_1 )| \le M_2.
\end{align}

By substituting (\ref{M})-(\ref{MM}) into (\ref{e}), the bilinear interpolation error $E_{sd}$ is expressed as 
\begin{align}
	\nonumber
	& E_{sd} = R(n,m) \le \frac{M_0}{2} (m-m_0) (m_1-m)  \\  \label{Esd}
	&+ \frac{[M_1(m_1-m)+M_2(m-m_{0})](n-n_0) (n_1-n)}{2(m_{1}-m_{0})}.
\end{align}

\bibliography{ref}

\end{document}